\documentclass[aps,footinbib,twocolumn,pra,superscriptaddress,groupedaddress]{revtex4}

\usepackage[utf8]{inputenc}
\usepackage[T1]{fontenc}

\usepackage{amssymb} \usepackage{color,graphicx} \usepackage{amsmath}
\usepackage{amsbsy} \usepackage{amsthm} 
\usepackage{bm} \usepackage{float} 
\usepackage{placeins,cals}
\usepackage{setspace,ulem}
\usepackage{physics,comment}
\usepackage{lineno}

\newcommand{\D}{\operatorname{d}}

\usepackage{braket}

\newcommand\rC{r_\text{\tiny C}}

\newcommand\x{{\bf x}}
\newcommand\y{{\bf y}}

\begin{document}

\title{Spontaneous collapse models lead to the emergence of classicality of the Universe}

\author{José Luis Gaona-Reyes}
\affiliation{Department of Physics, University of Trieste, Strada Costiera 11, 34151 Trieste, Italy}
\affiliation{Istituto Nazionale di Fisica Nucleare, Trieste Section, Via Valerio 2, 34127 Trieste, Italy}
\author{Lucía Menéndez-Pidal}
\affiliation{Departamento de Física Teórica, Universidad Complutense de Madrid, Parque de Ciencias 1, 28040 Madrid, Spain}
\author{Mir Faizal}
\affiliation{Irving K. Barber School of Arts and Sciences,
University of British Columbia Okanagan, Kelowna, BC V1V 1V7, Canada}
\affiliation{Canadian Quantum Research Center, 204-3002,
32 Ave Vernon, BC V1T 2L7 Canada}
\affiliation{CERN, Theoretical Physics Department, CH-1211 Geneva 23, Switzerland}
\author{Matteo Carlesso} 
\email{matteo.carlesso@units.it}
\affiliation{Department of Physics, University of Trieste, Strada Costiera 11, 34151 Trieste, Italy}
\affiliation{Istituto Nazionale di Fisica Nucleare, Trieste Section, Via Valerio 2, 34127 Trieste, Italy}
\affiliation{Centre for Quantum Materials and Technologies,
School of Mathematics and Physics, Queens University, Belfast BT7 1NN, United Kingdom}

\begin{abstract}
Assuming that Quantum Mechanics is universal and that it can be applied over all scales, then the Universe is allowed to be in a quantum superposition of states, where each of them can correspond to a different space-time geometry. How can one then describe the emergence of the classical, well-defined geometry that we observe? Considering that the decoherence-driven quantum-to-classical transition relies on external physical entities, this process cannot account for the emergence of the classical behaviour of the Universe. Here, we show how models of spontaneous collapse of the wavefunction can offer a viable mechanism for explaining such an emergence. We apply it to a simple General Relativity dynamical model for gravity and a  perfect fluid. We show that, by starting from a general quantum superposition of different geometries, the collapse dynamics leads to a single geometry, thus providing a possible mechanism for the quantum-to-classical transition of the Universe. Similarly, when applying our dynamics to the physically-equivalent Parametrised Unimodular gravity model, we obtain a collapse on the basis of the cosmological constant, where eventually one precise value is selected, thus providing also a viable explanation for the cosmological constant problem. Our formalism can be easily applied to  other quantum cosmological models {where we can choose a well-defined clock variable}.
 
\end{abstract}

\maketitle

\section{Introduction}
The common understanding is that Quantum Mechanics (QM) is a universal theory, being able to describe phenomena at all scales (from micro to macroscale). In this view, classical mechanics is just a macroscopic limit of QM, where a system's dynamics becomes classical due to the interaction with an \textit{external environment}. The latter acts as a measurer and makes the system \textit{decohere} (it loses quantum coherences) and the classical dynamics is  restored. However, the environment is also made of quantum particles and it should obey the rules of QM, thus making clear that we face a fundamental conundrum. John Bell stated this problem in simple terms \cite{bellagainst}:
 ``What exactly qualifies some physical systems to play the role of \textit{measurer}?''. Indeed, there is no prescription in where to draw a line between the measurers (also made of microscopic, quantum-mechanical particles) and the system being measured; such a division is purely arbitrary \cite{weinbergPRA}. This is the basis of the well-known --- for all practical purposes neglected --- \textit{quantum measurement problem} and, consequently, the \textit{quantum-to-classical transition}, which are  subjects of an active and growing research field \cite{AdlerbassiExact,carlessoNatPhys,Schlosshauer2005,Zurek2009,Jeong2014}.
The problem is {exacerbated} in the context of cosmology \cite{Kiefer1998}: if quantum theory is universal, the Universe should also be quantum. In particular, not only matter fields should be quantum, but also space-time itself, although what we observe is fully in line with classical General Relativity (GR).
To make an example,  the  fluctuations of the field of the Cosmic Microwave Background (CMB), whose origin -- according to frameworks such as inflationary cosmology \cite{Halliwell1987} or with bouncing scenarios \cite{Ashtekar2016} -- are quantum,  still deceive measurements. 
Here, we focus on the quantumness of space-time, rather than that of matter fields. 
Now, assuming that at its beginning the Universe is quantum and knowing that all what we observe is in line with classical predictions, {then a quantum-to-classical transition must have taken place before the CMB photon emission, which is the oldest observable signal in the Universe {and it can be described using classical matter fields on a classical geometry.}} 
Moreover, conversely to other systems, the Universe has no external physical entity that can act as a measurer, and thus the collapse due to a measurement cannot be the mechanism leading to its quantum-to-classical transition \footnote[1]{We must mention that there are models, such as  brane cosmology \cite{brax2004brane}, where an external physical entity is considered.}. For this reason, the Copenhagen interpretation of QM is not appropriate to describe such a transition. Indeed,  one would still require the interaction with an external  entity (for example an environment) that measures the system, and that is then averaged out.

Here, we focus on a solution that can be more suitable and comes from models of spontaneous wavefunction collapse \cite{bassighirardi, Bassi2013} (or simply collapse models). These {models} consistently describe the breakdown of quantum properties of a system, through the collapse of its wavefunction. 
Such a collapse is a scale-dependent phenomenon (the larger the system, the stronger the collapse), and it is implemented via non-linear and stochastic modifications of the Schr\"odinger equation. Notably, it is the fundamental dynamical equation of Nature which is modified to embed the collapse mechanism, and there is no need of external entities to make the system decohere or collapse.
Below, we will describe an application of collapse models in the field of quantum cosmology. We will show how such models can lead to the emergence of a classical Universe with a well-defined space-time geometry starting from a quantum Universe, whose  state can be in a  superposition of various geometries. Our approach is well different from previous ones, which focused on the quantum-to-classical transition of fields evolving on a well-defined classical background geometry. While our model {does not predict deviations from a classical model of cosmology after the CMB photon emission,} it allows for bridging the gap between a quantum theory of cosmology for earlier stages and a fully classical treatment for later ones.

Before dwelling in our model, we introduce the basis of our analysis. We use the $(-,+,+,+)$ convention for the metric, and we set the speed of light $c=1$ and $\kappa=8\pi G=1$.
For the sake of simplicity, we restrict the discussion to a flat Friedmann-Lemaître-Robertson-Walker (FLRW) Universe being maximally symmetric, homogeneous and isotropic (satisfying the cosmological principle). The corresponding metric is described by $\D s^2=-N^2(\tau)\D \tau^2 +a^2(\tau)h_{ij}\D x^i \D x^j$, where $h_{ij}$ is a flat Euclidean 3-metric called the spatial metric, and $a$ is the scale factor. Here, the lapse function $N$ represents the freedom of choosing a different time $(\tau)$ coordinate: for example, $N=1$ corresponds to the standard cosmological time, whereas $N=a^2$ to the  conformal time. The corresponding topology is $\mathcal{M}=\mathbb{R}\times \Sigma$, where $\Sigma$ is the flat spatial manifold. {Thus, the space-time is foliated in a collection of hypersurfaces connected by a time-like direction \cite{Kiefer2004}.}
To avoid divergences, we restrict  to finite  volumes $\int_\Sigma \D^3 x\, \sqrt{h}=V_0<\infty$, where $\Sigma$ is usually assumed to be a flat three dimensional torus. In such a way the metric is fixed.

\section{General Relativity with perfect fluid} In the context of GR, we will consider the dynamical evolution of gravity and a matter field, where the latter is modelled as a perfect fluid. Their total action reads
\begin{equation} 
    \mathcal{S}_\text{GR}=V_0 \int_\mathbb{R} \D \tau \left(-\frac{3\dot{a}^2a}{N}-N\frac{m}{a^{3w}}+m\dot{\chi} \right),
    \label{SGRFLRW}
\end{equation}
where $w$ defines the equation of state of the perfect fluid $p=w\rho$, with $p$ and $\rho(n)=\rho_0n^{1+w}$ being respectively the pressure and the density of the fluid and $n$ the particle number density. Here, $m$ corresponds to the energy density and relates to $n$ via $n a^3=(m/\rho_0)^{1/(1+w)}$, and $\chi$ is a Lagrangian multiplier being the conjugate coordinate to $m$. One can write Eq.~\eqref{SGRFLRW} in the Hamiltonian formalism, with the corresponding Hamiltonian reading 
\begin{table}[t]
    \centering
\caption{Change of variables to derive Eq.~\eqref{eq.Hamiltonian}. Left column: GR canonical transformation; Right column PUM transformation. Here we also use $\pi_a=-6V_0\dot a a/N$. }
    \label{tab1}
    \begin{tabular}{l|l}
    \toprule
GR canonical transf.& PUM transf.\\
\hline
$v=4\sqrt{{V_0}/{3}}{a^{{3(1-w)}/{2}}}/{(1-w)}$ & $v=2\sqrt{{V_0}/{3}}a^3$\\
$\pi_v=\pi_a a^{{(3w-1)}/{2}}/\sqrt{{12V_0}}$& $\pi_v=\pi_a/(a^2\sqrt{{12V_0}})$\\
    $t= {\chi}/{V_0}$& $t={T'}/{V_0}$ \\
    $\lambda=mV_0$& $\lambda=\Lambda V_0$\\
    \hline
    \end{tabular}
\end{table}
\begin{equation}\label{eq.Hamiltonian}
        \mathcal{H}=\bar{N}\left(-\pi_v^2+\lambda\right),
\end{equation}
where $\bar N=N a^{-3w}$ and one applies the GR canonical transformation presented in Tab.~\ref{tab1}.

As already mentioned, the lapse function $N$ (and thus $\bar N$) encodes the time reparametrisation freedom of our model (for example, one can take $N=1$ or $N=a^2$). However, since GR is a diffeomorphism invariant theory, i.e. it is invariant under changes of coordinates,
choosing an expression for $N$ merely corresponds to choosing a gauge.  Then, $N$ is a free parameter of our theory and enters Eq.~\eqref{SGRFLRW} as a Lagrange multiplier, which leads to the following 
\begin{equation}
    \mathcal{C}=-\pi_v^2+\lambda=0\, .
    \label{classconst}
\end{equation}
This is the  classical Hamiltonian constraint, which is also the starting point for our discussion. Both $\lambda$ and $\pi_v$ are constants of motion of the classical trajectories, and whose values are related to each other via Eq.~\eqref{classconst} which imposes
$\lambda=\pi_v^2$. In turn, this implies that $\lambda$ cannot take negative values, in full agreement with having $\lambda$  proportional to the energy density of the perfect fluid. {We remark that the Hamiltonian constraint in Eq.~\eqref{classconst} does not contain any time parameter.}

\section{Parametrised Unimodular Gravity} The  same expressions for the Hamiltonian in Eq.~\eqref{eq.Hamiltonian} and the classical constraint in Eq.~\eqref{classconst} can be derived in the context of {Parametrised} Unimodular gravity (PUM), which has the same metric and an action reading 
\begin{equation}
    \mathcal{S}_\text{PUM}= \int_{\mathbb{R}} \D \tau \left(\pi_a \dot{a} +\Lambda \dot{T}'-N\left[ -\frac{1}{12}\frac{ \pi^2_a}{V_0 a}+V_0 a^3\Lambda\right] \right) ,
    \label{SPUMmodel}
\end{equation}
where $\pi_a$ is the conjugate momentum to $a$, $T'=T/V_0$ with $T$ defines $\sqrt{-g}=\partial_a T^a$ and $\Lambda$ is a dynamical field whose equation of motion is $\partial_a \Lambda=0$. {Physically, PUM 
is a theory of gravity in which, instead of having diffeomorphism invariance as in GR, we restrict ourselves to transformations that leave the determinant of the metric $\sqrt{-g}$ fixed \cite{Unruh1989a,Carballo-Rubio2022}. It is possible to recover full diffeomorphism invariance with the addition of extra fields \cite{henn1989}}, thus the cosmological constant $\Lambda$ appears very naturally as a constant of motion ($\partial_a \Lambda=0$) with its canonically conjugate coordinate being $T'$. Classically, there is no quantitative difference between GR+perfect fluid and PUM, except from the different treatment of the cosmological constant. 
In the former $\Lambda$ is a fixed constant of Nature, whereas in the latter  is a canonical variable. At the quantum level, PUM addresses the so-called problem of time (see below and in SM) of Quantum Gravity by having introduced a natural clock variable.

A common point of these two models is that we have two canonically conjugated pairs of variables
\begin{equation}\label{conjugatevar}
    \lbrace v, \pi_v \rbrace=1,\quad \lbrace t, \lambda \rbrace=1.
\end{equation}
The physical interpretation of the first pair is {common to both models}, $v$ is related to the scale factor $a$ and $\pi_v$ is its conjugate momentum.
The physical interpretation of the second pair instead depends on the model. In GR, $\lambda$ is related to the perfect fluid degree of freedom, while in PUM it encodes the unimodular cosmological constant.
Despite coming from two different points of view, the mathematical structure of both descriptions (GR+perfect fluid and PUM) is equivalent. In the following, $\lambda$ may refer to the perfect fluid energy density or the cosmological constant, however, none of the results is affected by the interpretation of this variable.

Now, we proceed to quantise the model. We upgrade the classical constraint $\mathcal C$ in Eq.~\eqref{classconst} to a quantum operator $\hat{\mathcal{C}}$ that will be applied to the wavefunction $\ket{\Psi}$ of the Universe and gives $\hat{\mathcal{C}}\ket{\Psi}=0$. Namely, in the $(v,t)$ representation this becomes the Wheeler-DeWitt (WDW) equation, which reads
\begin{equation}\label{eq.WDW}
    \left( \frac{\partial^2}{\partial v^2}-i \frac{\partial}{\partial t}\right)\Psi(v,t)=0,
\end{equation}
where the partial derivatives are obtained with the mapping $\hat \pi_v\to-i {\partial}/{\partial v}$ and $\hat \lambda\to-i {\partial}/{\partial t}$, and we set $\hbar=1$. {As neither $t$ nor $v$ are external parameters of the theory, Eq.~\eqref{eq.WDW} is timeless. This is the so-called \textit{problem of time} in quantum gravity \cite{Anderson2012}.} 
{To address this issue, one considers an internal dynamical variable as \textit{relational clock}, and thus allows the Universe to have a non-trivial dynamics. Therefore, by identifying $t$ as the temporal variable {--- the choice of a clock is a fundamental point of our approach ---} Eq.~\eqref{eq.WDW} has the same structure of a Schr\"odinger equation with effective Hamiltonian $\hat H=-\hat \pi_v^2$}. 
The same clock choice has been considered in also previously \cite{Ali2018}.

\section{The model} The fundamental point for our discussion is that the wavefunction of the Universe  is governed by the WDW equation \eqref{eq.WDW}, which allows and preserves superpositions in the four variables displayed in Eq.~\eqref{conjugatevar}. Indeed, the WDW equation is linear and there are no external factors that can decohere or collapse the wavefunction. In particular, the Universe can be in a  superposition of different values of $\lambda$, which is related to the energy density $m$ (for GR) or to the cosmological constant $\Lambda$ (for PUM), and thus of different space-time geometries. 
\begin{figure*}
    \centering
    \includegraphics[width=0.8\linewidth]{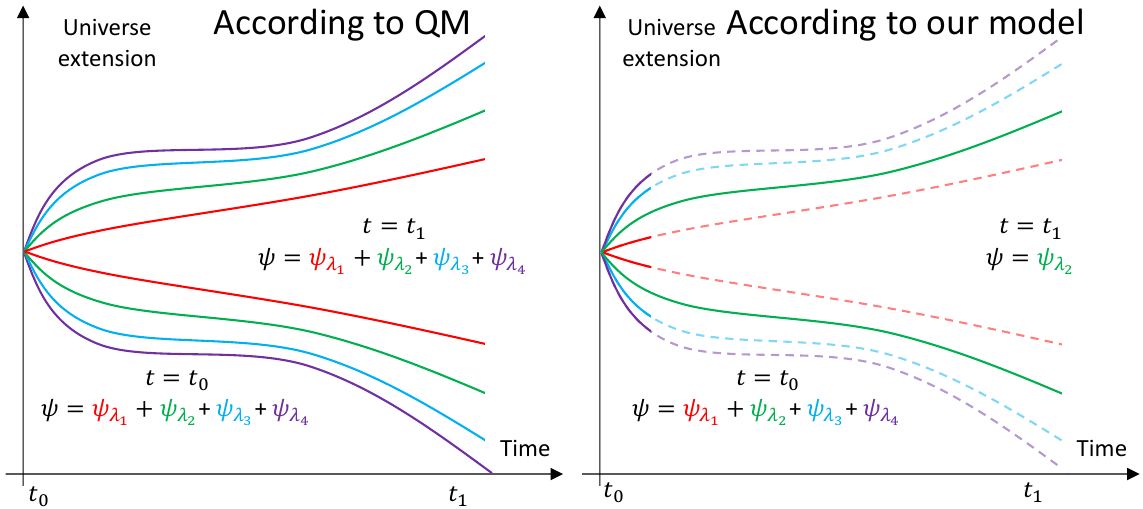}
    \caption{(Left panel) According to quantum mechanics, 
    an initial superposition of the Universe $\Psi=\Psi_{\lambda_1}+\Psi_{\lambda_2}+\Psi_{\lambda_3}+\Psi_{\lambda_4}$ at time $t_0$  is conserved under the unitary dynamics of the Schr\"odinger equation. The state of the Universe never collapses in one eigeinstate of $\hat \lambda$ for any later time $t_1$.
 (Right panel) The introduction of collapse terms in the Wheeler--DeWitt equation allows for the collapse of the wavefunction from the initial superposition at time $t_0$ into a single eigenstate of $\hat{\lambda}$ (e.g., $\Psi=\Psi_{\lambda_2}$) at a time $t_1$. }
    \label{fig:enter-label}
\end{figure*}

Now we will show that, by suitably modifying the WDW equation and including the collapse terms, we are able to make the Universe collapse in a specific eigeinstate of $\hat \lambda$ corresponding to a given space-time geometry. 
Following the standard prescriptions of collapse models, which are  reported in the Methods, we modify the WDW equation \eqref{eq.WDW} by adding stochastic and non-linear terms. This  gives
\begin{widetext}
\begin{equation} \label{modwdw}
\left[\hat{{H}}+i\left(\hat{{A}}-\braket{\hat{{A}}}_t\right)\frac{\D W_t}{\D t}-\frac{i}{2}\left(\hat{{A}}-\braket{\hat{{A}}}_t\right)^2 -i  \frac{\partial}{\partial t} \right]\Psi(v,t)=0,
\end{equation}
\end{widetext}
where $\hat{{A}}$ is the yet to be chosen collapse operator, $\braket{\hat{{A}}}_t~=~\braket{\Psi_t|\hat A|\Psi_t}$, and $W_t$ represents a Wiener process in $t$.
How to fix the collapse operator $\hat A$ depends on the basis  {in which} one wants to observe the collapse of the wavefunction. The standard choice in non-relativistic collapse models such as  the Continuous Spontaneous Localisation (CSL) \cite{CSL1,CSL2} or the Diósi-Penrose (DP) \cite{DP1,DP2} models falls on the mass density, i.e. a function of the position operator $\hat x$, so that eventually macroscopic systems have a well-localised position as described by classical mechanics. A more complex situation arises when moving to a relativistic framework \cite{jones}, although some proposals
have been suggested \cite{Bedingham}. Here, we consider the choice of $\hat A= \epsilon \hat H$, where $\epsilon$ encodes the rate of collapse and quantifies the coupling between the collapse noise and {the wavefunction of the Universe}. Finally,  $\hat H$ is its Hamiltonian, being a natural (although not unique) relativistic generalisation of the non-relativistic mass density. 
In particular, $\hat H$ is related to the operator $\hat \lambda$ via the quantum version of the classical constraint in Eq.~\eqref{classconst}, i.e.~$\hat H=-\hat \pi_v^2=-\hat\lambda$. This implies that such a choice for the collapse operator $\hat A$ imposes a localisation in $\lambda$. 
From Eq.~\eqref{modwdw} we can derive the dynamics of the mean $\braket{\hat \lambda}_t$ and its variance $\sigma^2_{\lambda,t}=\braket{\hat \lambda^2}_t-\braket{\hat \lambda}_t^2$, which respectively read
\begin{subequations}
\begin{align}
        &\D\braket{\hat \lambda}_t=-2\epsilon \sigma^2_{\lambda,t}\D W_t,\\
        \label{eq.sigma}&\D\sigma^2_{\lambda,t}=-4\epsilon^2(\sigma_{\lambda,t}^2)^2 \D t-2\epsilon \Sigma^{(3)}_t\D W_t,
\end{align}
\end{subequations}
where $\Sigma^{(3)}_t=(\braket{\hat \lambda^3}_t-3\braket{\hat \lambda}_t\braket{\hat \lambda^2}_t+2\braket{\hat \lambda}_t^3)$.
Notably, the solution for the mean critically depends on the history of the variance $\sigma^2_{\lambda,t}$, namely $\braket{\hat \lambda}_t=\braket{\hat \lambda}_{t_0}-2\epsilon\int_{t_0}^t \sigma^2_{\lambda,s}\D W_s$. Thus, when the collapse process localises the wavefunction in $\lambda$, i.e.~after a time $t^*$ we have  $\sigma^2_{\lambda,t>t^*}\sim0$, the corresponding mean becomes fixed and does not change $\braket{\hat \lambda}_{t>t^*}= \braket{\hat \lambda}_\infty$. Physically, the collapse dynamics in Eq.~\eqref{modwdw} is driving an arbitrary initial state $\Psi$, which can also be in a superposition of $\lambda$ with $\sigma^2_{\lambda,t_0}\neq0$, into a eigenstate of the operator $\hat \lambda$ with $\sigma^2_{\lambda,t>t^*}\sim0$. Thus, at the end of the collapse process, one obtains a Universe with a well-defined space-time geometry with a fixed value for $m$ or $\Lambda$ respectively in the case of GR or PUM. Notably, such a localisation is  triggered for any realisation of the noise $W_t$, and one does not need to average over different realisations to observe the effect. A schematic representation of the effects of the introduction of the collapse terms in the WDW equation is shown in Fig.\ref{fig:enter-label}.

\begin{figure}[t]
    \centering
\includegraphics[width=\linewidth]{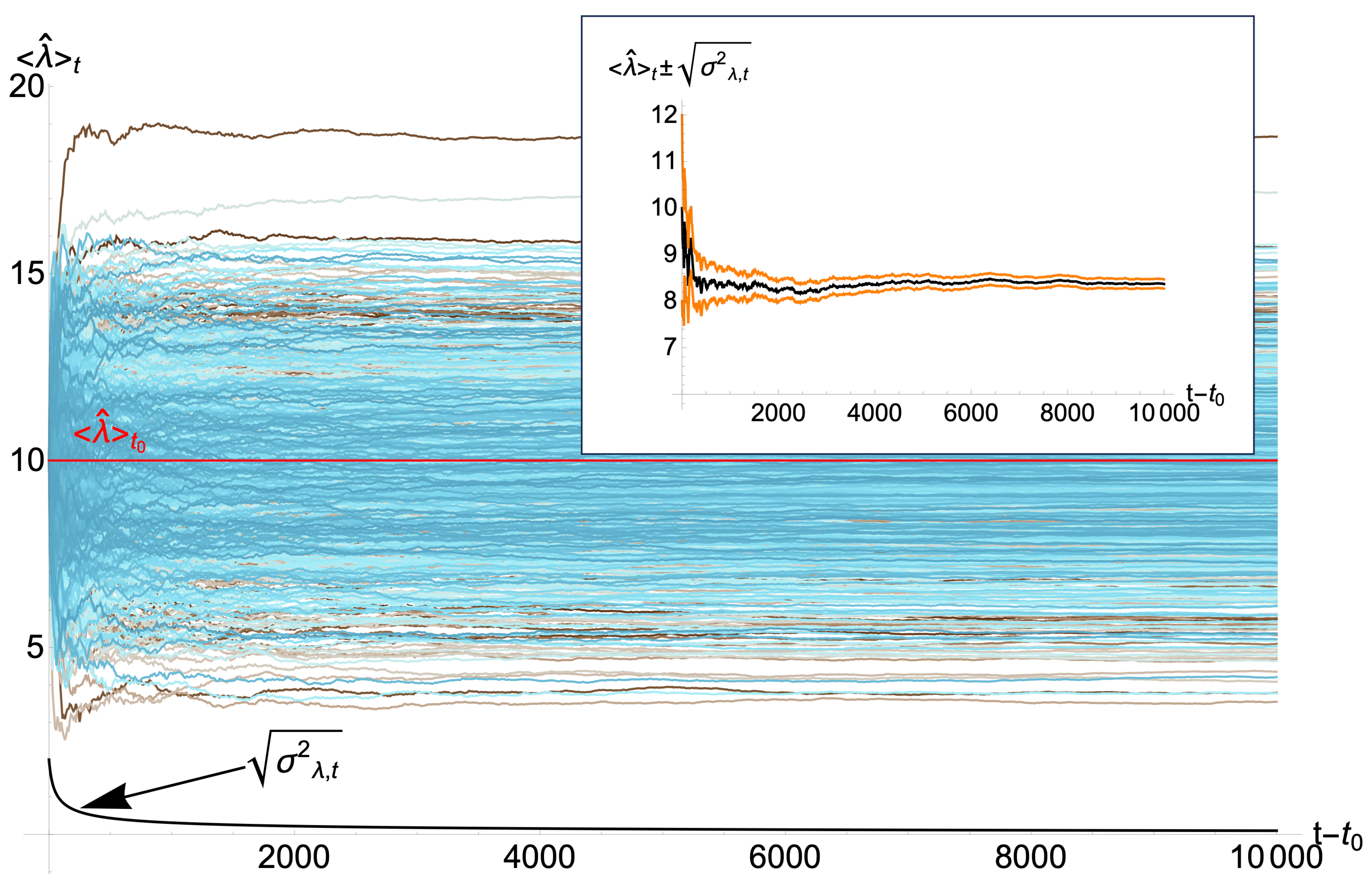}
    \caption{Mean $\braket{\hat \lambda}_t$ for 1000 different realisations of the noise field $W_t$, with $\epsilon=0.05$ and an initial state with $Q(\lambda)\propto e^{-(\lambda-\lambda_0)^2/4\sigma_0^2}$ with $\lambda_0=10$ and $\sigma_0^2=4$. The red line highlights the initial value of $\braket{\hat \lambda}_{t_0}$, while the black line shows the evolution of $\sigma_{\lambda,t}$ under the Gaussian assumption. The inset show a single realisation of the noise: in black $\braket{\hat \lambda}_t$, in orange $\braket{\hat \lambda}_t\pm \sqrt{\sigma^2_{\lambda,t}}$. 
    }
    \label{fig.gauss}
\end{figure}

The dynamics of the variance can be analytically solved under the assumption of the Gaussianity of the state. In such a case, the second term in Eq.~\eqref{eq.sigma} vanishes (i.e., $\Sigma^{(3)}_t=0$) and we find $\sigma^2_{\lambda,t}=\sigma^2_{\lambda,t_0}/(1+4\epsilon^2(t-t_0)\sigma^2_{\lambda,t_0})$.  To {provide}
an explicit example, in Fig.~\ref{fig.gauss} we show the evolution of the mean $\braket{\hat\lambda}_t$ for different realisations of the noise $W_t$ starting from an
initial  Gaussian state $\ket{\Psi}_{t_0}=
\int_0^{\infty}\D \lambda\, Q(\lambda)\ket{\lambda}$ such that $Q(\lambda)\propto e^{-(\lambda-\lambda_0)^2/4\sigma_0^2}$. From the numerical simulations we can notice that $\braket{\hat \lambda}_t$ does not fully localise to a fixed value within the time-scale of the simulation. This is due to the slow decay of $\sigma_{\lambda,t}^2\sim (t-t_0)^{-1}$, which implies that the localisation process runs on a long time-scale.
Beyond the Gaussian regime, one needs to employ perturbative approaches. These are discussed in the Appendix.

\section{Discussion} Our model describes how a classical, well-defined space-time geometry (a localised state in $\lambda$, that can be related to a classical Universe) can naturally emerge from any initial state, such as a quantum superposition in $\lambda$. {Moreover, we remark that this value of $\lambda$ in general differs from the quantum expectation value of $\hat{\lambda}$ at an initial time.}
 Thus, the inclusion of collapse terms into the WDW equation offers a suitable mechanism for the description of the quantum-to-classical transition in the cosmological context.
 This mechanism does not invoke nor require the presence of an external physical entity acting as a \textit{measurer} to collapse the Universe's state. 
 We remark that by choosing the collapse operator as being proportional to the Hamiltonian of the system allows a wide application of the model. This comprises non-relativistic scenarios, where its action reduces to that of the CSL model, as well as configurations where matter fields can be included.
Even though the WDW equation can be considered as a specific toy model, we want to stress that our main result does not depend on the details of the model we used. Other quantum cosmology models, where for example one considers different definitions of the time variable, will provide similar results.
We underline that our approach for the quantum-to-classical transition in cosmology differs
from previous proposals \cite{Okon2014,Banerjee2017}. There, the aim was to tackle the problem of time and they do it by introducing a collapse dynamics with respect to the gauge coordinate $\tau$. Conversely, we introduce a new dynamical variable $t$ (not a mere gauge coordinate), which can be used to determine a physical dynamics of the system and with respect to which a quantum-to-classical transition is governed. In  SM, we provide further comparison of our approach with respect to the existing literature.

We underline that Eq.~\eqref{modwdw} does not necessarily solve the problem of time in Quantum Gravity. Indeed, we are already assuming the variable $t$ to play the distinguished role of clock and we are building the collapse dynamics over that structure. In such a way collapse terms
cannot come as a solution to the problem of time, but are just related to the clock choice. {Nevertheless, we underline that having made a clock choice is fundamental to make any claim on the collapse having taken place or not. It is not clear if  a collapse dynamics can be implemented in a context where there is no clock variable. We leave this for future research.}

{An interesting aspect of our approach, which comes as a byproduct of the model, is that it provides an alternative explanation for the cosmological constant problem. Conventionally, a classical ensemble of Universes in a multiverse, each with a different cosmological constant, is assumed to exist, and then the anthropic principle explains why we observe a specific value of the cosmological constant. Here, we do not need such a construct. We just have a quantum superposition of different cosmological constants at very early stages of the Universe, which later collapses to the observed value of the cosmological constant.}

Notably, there is an important difference between the studied context and a typical experimental situation. In the latter, one can repeat the experiment several times, where at each realisation the collapse noise acts differently and makes the system's wavefunction collapse on a possibly different eigenvalue of the collapse operator. Averaging over these realisations, one can compute the predicted deviations of collapse model with respect to QM. Conversely, in the context of cosmology, the experiment --- being the Universe --- can be ran only once. 
Then, one has a unique realisation of the collapse noise, which leads to the collapse to a specific geometry, and observables (such as the cosmological constant $\Lambda$) take specific values. Therefore, one cannot distinguish a quantum Universe with a collapse as described by our model from a classical Universe. In general, to confirm or falsify the existence of collapse models, one will have to restrain to repeatable experiments, which is not the focus of our work. However, in principle one could restrict the possible values of the collapse rate $\epsilon$ by requiring {the emergence of} a well-defined geometry before the CMB photon emission. 

Finally, we underline that our results are very general and would hold for any valid approach to quantum cosmology where a suitable clock is defined, not only GR+perfect fluid or PUM. Examples are models where the dust time, the scale factor, the axion field  or the dilaton field are used as a clock. Our approach can also be generalised  to Loop Quantum Cosmology, where the clock is usually the scalar matter field.

\section*{Acknowledgments}
We thank Angelo Bassi for providing helpful comments on an early draft of the work.
LMP would like to thank Rita Neves for useful discussions on the topic.
JLGR and MC acknowledge the EIC Pathfinder project QuCoM (GA No.~101046973). LMP is supported by the Leverhulme Trust. 
 MF acknowledges the support of BRIN.  
MC is supported by UK
EPSRC (Grant No.~EP/T028106/1) and PNRR PE National Quantum Science and Technology Institute (PE0000023).  

   \onecolumngrid
\appendix

\subsection*{Solution of the WDW equation in GR and PUM}

The solution to the WDW equation shown in Eq.~(6) of the main text can be found via separation of variables and yields the general form
\begin{equation}
    \Psi(v,t)=\int_{-\infty}^{\infty}\D k\, e^{i k^2 t }\left( A(k)e^{i k v}+B(k)e^{-ikv} \right)  +\int_{-\infty}^{\infty}\D \kappa\, e^{-i \kappa^2 t }\left( C(\kappa)e^{ \kappa v}+D(\kappa)e^{-\kappa v} \right),
\end{equation}
where $A(k)$, $B(k)$, $C(\kappa)$, and $D(\kappa)$ are functions to be determined. Positive values of $k$ may be directly associated to $\sqrt{\lambda}$. To make such states evolve, one needs a notion of time. In the absence of an absolute external parameter, both dynamical fields $v$ and $t$ are equally good candidates to play this role. Here, we choose $t$ as the time variable so that the WDW equation can be seen as a Schrödinger equation. 
Nevertheless, we want to stress that $v$ is an equally valid choice. This embarrassment of riches is known as the \textit{problem of choice}. It has been seen that different clock choices lead to different  descriptions of the quantum theory. For example, by choosing $t$ as the time one obtains the Schrödinger equation, while choosing $v$ leads to the Klein-Gordon equation 
 \cite{Gielen2020}. In this work, we focus  in how to obtain a classical universe from a quantum wavefunction rather than trying to find a definitive solution to this issue.

The clock choice comes with an inner product --- or equivalently a Hilbert space --- specification. In our case, as the WDW equation is a Schrödinger equation in $t$, the inner product is defined as
\begin{equation}
 \braket{\Phi|\Psi}=\int_0^\infty \D v \,  {\Phi^*}(v)\Psi(v),
\end{equation}
where ${\Phi}^*$ denotes the complex conjugate of $\Phi$. Note that the $v$ integration range goes from 0 to $\infty$. Indeed, classically, $v$ is only defined to be non-negative (it is a positive power of the scale factor). Before introducing the collapse terms, we want our wavefunctions to evolve unitarily according to QM, meaning  that the Hamiltonian term $\partial^2/\partial v^2$ must be self-adjoint. To ensure that, one must impose the following additional condition \cite{Al-Hashimi2021} 
\begin{equation} \label{eq.bcond}
    \gamma \Psi(0,t)-\pdv{}{v}\Psi(0,t)=0,
\end{equation}
where $\gamma\in \mathbb{R}\cup\lbrace \infty \rbrace$ is a free parameter. The condition in Eq.~\eqref{eq.bcond} can be seen as a reflection around $v=0$, whose requirement is motivated as follows. We want our solutions to have a well preserved norm, but $v$ is only defined on the half-line. Thus, we must ensure that there is no probability flow at $v=0$, and Eq.~\eqref{eq.bcond} guarantees that this is the case. The parameter $\gamma$ represents the freedom in how states are reflected. In a nutshell, the only solutions of Eq.~(6) of the main text that evolve unitarly (when no collapse is considered) are those satisfying Eq.~\eqref{eq.bcond}.
 Their explicit form reads
 \begin{equation}\label{wavefunction}
    \Psi(v,t)=\int_{0}^\infty \frac{\D \lambda}{\sqrt{2\pi}}\,  Q(\lambda)e^{i\lambda t}\psi_{\lambda, \gamma}(v)\, ,
\end{equation}
where
\begin{equation}
      \psi_{\lambda, \gamma}(v)=\frac{1}{\sqrt{2\sqrt{\lambda}}}\left( e^{-i\sqrt{\lambda}v}+\frac{i\sqrt{\lambda}+\gamma}{i \sqrt{\lambda} -\gamma}e^{i\sqrt{\lambda}v} \right).
\end{equation}
Here, $\lambda$ is assumed to be positive or zero and the functions $\psi_{\lambda,\gamma}(v)$ are Dirac delta normalised: $\int_0^\infty \D v \psi^*_{\lambda_1, \gamma}(v)\psi_{\lambda_2, \gamma}(v)=2\pi \delta(\lambda_1-\lambda_2)$, while {the weight} $Q(\lambda)$ characterises the superposition of different values of $\lambda$, with $\int_0^\infty \D \lambda \abs{Q(\lambda)}^2=1$. 

Note that, in order to obtain numerical results, the value of $\gamma$ needs to be fixed, where  each value of $\gamma$ corresponds to a specific reflection around $v=0$.  Whilst different choices of $\gamma$ imply different eigenstates $\psi_{\lambda,\gamma}=\braket{v|\lambda}$, the addition of collapse terms to the WDW equation always ensures that the Universe transitions from a superposition in $\lambda$ to a localised state. So, this would not have altered the conclusions of our approach.  In this work, we have taken the standard choice of $\gamma=\infty$ (another typical choice is $\gamma=0$), but any value of $\gamma$ would also be valid. Then, $\psi_{\lambda,\infty}(v)$ takes a very simple form, namely:
\begin{equation}
    \braket{v|\lambda}=\psi_{\lambda,\infty}(v)=-\frac{\sqrt{2}i}{\lambda^{1/4}}\sin(\sqrt{\lambda} v),
\end{equation}
which can be seen as a superposition of plane waves incoming to the classical singularity and outgoing from it.

Without the addition of collapse terms, the universe will never localise in $\lambda$. Indeed, given a generic state, one has
\begin{equation}
    \begin{aligned}
        \braket{\hat{\lambda}}_t&=\int_0^\infty \dd \lambda \abs{Q(\lambda)}^2\lambda\, , \\
        \sigma^2_{\lambda,t}&=\left[\int_0^\infty \dd \lambda \abs{Q(\lambda)}^2\lambda\right]^2-\int_0^\infty \dd \lambda \abs{Q(\lambda)}^2\lambda^2\, .
    \end{aligned}
\end{equation}
Both of these expressions are time independent; if the Universe starts with a certain expectation value $\braket{\hat \lambda}_0$ and spread $\sigma^2_{\lambda,0}$, it will remain that way throughout the evolution. In the main text, we focused in Gaussian states, $Q(\lambda)\propto e^{-(\lambda-\lambda_0)^2/4\sigma_0^2}$, which gives
\begin{equation}
\begin{aligned}
  &\braket{\hat{\lambda}}_t=\lambda_0+  \frac{\sqrt{\frac{2}{\pi }} \sigma_0e^{-\frac{\lambda_0^2}{2 \sigma_0^2}}}{\erf\left(\frac{\lambda_0}{\sqrt{2} \sigma_0}\right)+1},\\
&    \sigma^2_{\lambda,t}={\sigma_0}^2 \left(1-\frac{2 e^{-\frac{{\lambda_0}^2}{{\sigma_0}^2}}}{\pi  \left({\erf}\left(\frac{\lambda_0}{\sqrt{2} {\sigma_0}}\right)+1\right)^2}\right)+\frac{ \sqrt{\frac{2}{\pi }} {\lambda_0} {\sigma_0}e^{-\frac{{\lambda_0}^2}{2 {\sigma_0}^2}}}{{\operatorname{erfc}}\left(\frac{{\lambda_0}}{\sqrt{2} {\sigma_0}}\right)-2},
\end{aligned}
\end{equation}
which, we remark, are time-independent expressions.
Conversely, with the addition of collapse terms, we are not only able to make $\sigma^2_{\lambda,t}$ go to zero, but the value of $\lambda$ after the collapse can be different from the initial one, which corresponds to 
$\braket{\hat{\lambda}}_0$.

\subsection*{General structure of collapse models dynamics}

Collapse models are modifications of the standard quantum theory, where the Schr\"odinger equation is phenomenologically modified to include a spontaneous collapse of the wavefunction \cite{bassighirardi,Bassi2013,Arndt2014,carlessoNatPhys}. Such modifications involve non-linear and stochastic terms, whose structure --- under fairly general requirements --- is fixed. The requirements are \textit{i)} the occurrence of the collapse (or localisation) of the wavefunction in one of the eigenstates of a chosen collapse operator, with a probability following the Born rule, and \textit{ii)} not to violate the causality principle. The former requirement is provided by the non-linear structure of the collapse equation, which is necessary for having an actual collapse and not only a decoherence-like process (where the coherences are lost). The later requirement is satisfied by the stochastic nature of the modifications, which prevent superluminal signalling to occur. 
The general structure of the collapse models equation for the wavefunction $\ket{\Psi_t}$ reads
\begin{equation}
\begin{aligned}\label{Collapseeq2}
\D \ket{\Psi_t}=\left(-i \hat{H}\D t +\int \D \x \, \hat A_\text{N}(\x) \xi_t(\x) \D t-\frac{1}{2}\int \D \x \int\D \y\, G(\x,\y) \hat A_\text{N}(\x)\hat A_\text{N}(\y)\D t \right)\ket{\Psi_t},
\end{aligned}
\end{equation}
where  $\hat{H}$ is the Hamiltonian of the system, and $\hat A_\text{N}(\x)=(\hat{A}(\x)-\braket{\hat{A}(\x)}_t)$, with $\hat{A}(\x)$ being the collapse operator defined in space $\x$, and $\braket{\hat{A}(\x)}_t=\braket{\Psi_t|\hat A(\x)|\Psi_t}$. 
 The noise field $\xi_t(\x)$ has vanishing mean $\mathbb{E}[\xi_t(\x)]=0$, and correlation given by
\begin{equation}
\mathbb{E}[\xi_t(\x)\xi_{t'}(\x')]=G(\x,\y)\delta(t-t').
\end{equation}
Different choices of $\hat A(\x)$ and $G(\x,\y)$ determine different specific models, such as the Di\'osi-Penrose  \cite{DP1,DP2} (DP), the Continuous Spontaneous Localization  \cite{CSL1,CSL2} (CSL)  models (being the two most studied and tested collapse models), or  the model introduced  in the main text. Table \ref{tab2} summarises the specific choices for $\hat A(\x)$ and $G(\x-\y)$ required to construct these models. 
\begin{table}[t]
    \centering
    \caption{Choices for collapse operators $\hat{A}(\x)$ and $G(\x,\y)$ for the Di\'osi-Penrose (DP) \cite{DP1,DP2}, the Continuous Spontaneous Localization (CSL) \cite{CSL1,CSL2} and the model introduced in the main text. Here, $\hat M(\x)$ is the mass density operator, $\hat H$ is the Wheeler--DeWitt Hamiltonian; $\gamma$, $\rC$ and $\epsilon$ are free collapse parameters, while $G$ is the gravitational constant. 
    }
    \label{tab2}
    \begin{tabular}{c|c|c}
    \toprule
Model& $\hat{A}(\x)$ &  $G(\x,\y)$ \\
\hline 
CSL & ${\sqrt{\gamma}}\hat{M}(\x)/m_0$ & $ \exp\left[-{(\x-\y)^2}/{4 \rC^2}\right]/(4 \pi \rC^2)^{3/2}$  \\
DP & $\hat{M}(\x)$ & ${G}/{|\x-\y|}$  \\
Main & $\epsilon \hat H \delta(\x-\x_0)$&$1$ \\
    \hline
    \end{tabular}
\end{table}  

Another important feature for a good collapse model is to be endowed with an amplification mechanism, which ensures that the collapse action is negligible on microscopic systems (which are found in quantum superpositions), while becoming strong and effective as a growing function of the mass of the system. For this reason, one typically employs a collapse operator related to the mass of the system (for example, the mass density operator or the Hamiltonian operator --- as a relativistic extension of the mass density operator).

It is important to remark that, in contrast to the different interpretations of QM, collapse models are falsifiable theories. Indeed, there are qualitative differences between the predictions of collapse models and standard quantum theory, that can be used to 
test and constrain the phenomenological collapse parameters \cite{carlessoNatPhys}.

\subsection*{Perturbative approach for the dynamics}

Here, we show how to tackle Eq.~(7) of the main text via a perturbative approach.
We start by assuming the following perturbative ansatz for the solution $\ket{\Psi_t}$, expressed as a Taylor expansion up to the second order in $\epsilon$. This can be written in the $\{\ket\lambda\}$ basis as
\begin{equation}\label{eq.sol.pert}
\ket{\Psi_{t}}=\int_0^\infty \D \lambda\, Q(\lambda) (\sum_{j=0}^2 \epsilon^j{K}_{j,t}(\lambda)) \exp \left[{i}{}(t-t_0)\lambda \right]\ket{\lambda}, 
\end{equation}
where
$Q(\lambda)$ determines the  superposition of the initial state $\ket{\Psi_{t_0}}=\int_0^\infty\D\lambda\,Q(\lambda)\ket\lambda$ with the normalisation constraint reading $\int_0^\infty\D\lambda\,|Q(\lambda)|^2=1$. By inserting such an ansatz in Eq.~(7) of the main text,
we can derive the explicit expression of the $K_{j,t}(\lambda)$ functions. In particular, we find
 $K_{0,t}(\lambda)=1$, $K_{1,t}(\lambda)=(\braket{\hat \lambda}_{t_0}-\lambda)(W_t-W_{t_0})$ and 
\begin{equation}
\begin{aligned}
K_{2,t}(\lambda)=(t-t_0)\left[
\sigma_{t_0}^2(\hat \lambda)-(\lambda-\braket{\hat \lambda}_{t_0})^2
\right]+\tfrac{1}{2}(W_t-W_{t_0})^2\left[(\lambda-\braket{\hat \lambda}_{t_0})^2 - 2 \sigma_{t_0}^2(\hat \lambda)\right],
\end{aligned}
\end{equation}
where $\braket{\hat \lambda}_{t_0}$ and $\sigma_{\lambda,t_0}^2$ are respectively the mean and the variance of $\hat \lambda$ at time $t_0$. Once the explicit expressions for $K_{j,t}(\lambda)$ are plugged in Eq.~\eqref{eq.sol.pert}, we can derive the corresponding expressions for the mean $\braket{\hat \lambda}_{\text{per},t}$ and the variance $\sigma^2_{\text{per},t}$, where the label ``per'' indicates the perturbative derivation. They read
\begin{subequations}\label{meanvarnonpert}
\begin{equation}
    \begin{aligned}
        &\braket{\hat \lambda}_{\text{per},t}=\braket{\hat \lambda}_{t_0}-2\epsilon(W_t-W_{t_0})\sigma_{\lambda,t_0}^2-2\epsilon^2\left[(t-t_0)-(W_t-W_{t_0})^2\right]\braket{\hat \lambda[(\hat \lambda-\braket{\hat \lambda}_{t_0})^2-\sigma_{\lambda,t_0}]}_{t_0},
    \end{aligned}
\end{equation}
and
        \begin{equation}
    \begin{aligned}
      \sigma^2_{\text{per},t}&=\sigma^2_{\lambda,t_0}-2\epsilon(W_t-W_{t_0})\left[\braket{\hat \lambda^3}_{t_0}-\braket{\hat \lambda}_{t_0}(\braket{\hat \lambda^2}_{t_0}-2\sigma_{\lambda,t_0}^2)\right]-4\epsilon^2(W_t-W_{t_0})^2(\sigma_{\lambda,t_0}^2)^2\\
      &-2\epsilon^2\left[  (t-t_0)-(W_t-W_{t_0})^2 \right]\left[\braket{\hat \lambda^4}_{t_0}-4\braket{\hat \lambda}_{t_0}\left(\braket{\hat \lambda^3}_{t_0}+\braket{\hat \lambda}_{t_0}^3-2\braket{\hat \lambda^2}_{t_0}\braket{\hat \lambda}_{t_0}\right)-\braket{\hat \lambda^2}_{t_0}^2\right].
    \end{aligned}
\end{equation}  
\end{subequations}
Notably, these expressions depend on higher momentum expectation values $\braket{\hat \lambda^3}_{t_0}$ and $\braket{\hat \lambda^4}_{t_0}$, which can be simply computed given the initial state.

\begin{figure}[th]
    \centering
    \includegraphics[width=0.4\linewidth]{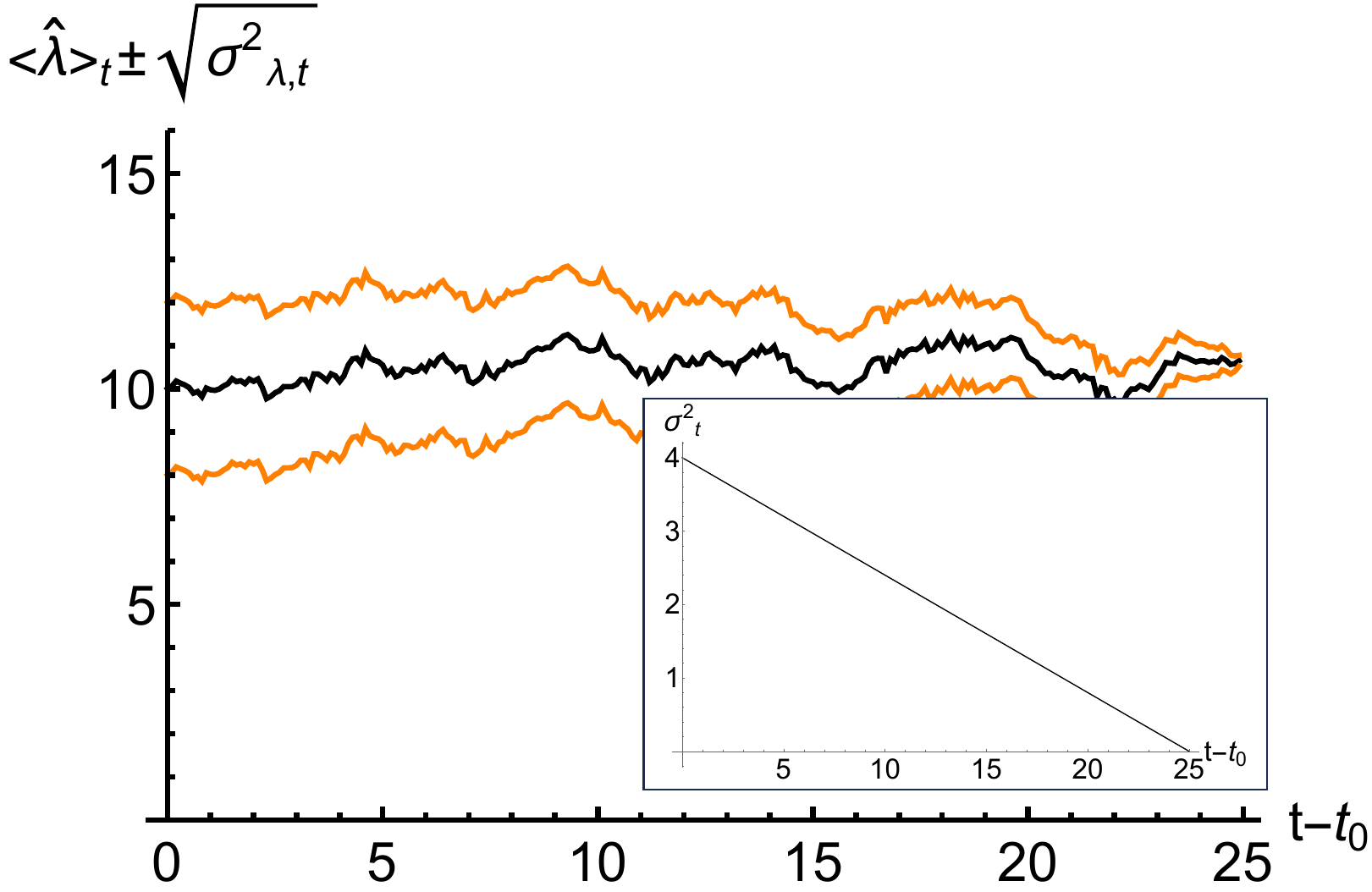}\includegraphics[width=0.4\linewidth]{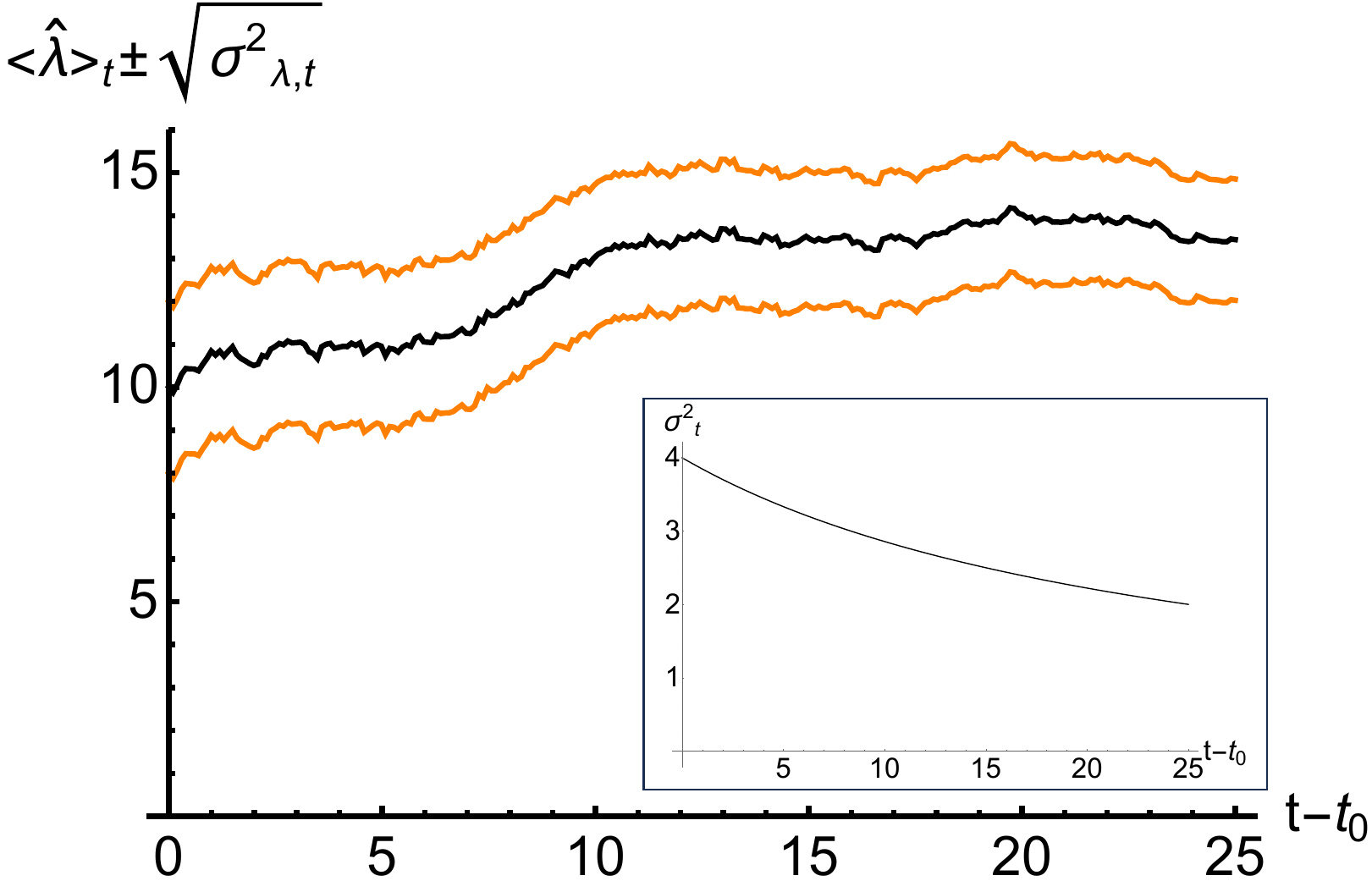}
    \caption{Comparison of the solutions $\braket{\hat \lambda}_t$ as expressed respectively in Eq.~\eqref{eq.lambda.pert.gauss} and Eq.~\eqref{eq.lambda.exact.gauss}. The insets report the corresponding variances $\sigma^2_t$ [cf.~Eq.~\eqref{eq.sigma.pert.gauss} and Eq.~\eqref{eq.sigma.exact.gauss}]. 
    The perturbative solution (left panel) is associated to a fast decaying variance, while the exact solution (right panel) is localised on a much longer time-scale.}
    \label{fig.comparison}
\end{figure}

To quantify the goodness 
of the perturbative approach, we consider the evolution under the assumption of an initial Gaussian distribution, which is taken to be equal as the one in the main text, i.e.~$Q(\lambda)\propto e^{-(\lambda-\lambda_0)^2/4\sigma_0^2}$.
This allows for a one-to-one comparison with the exact solution reported in the main text.
In particular, since the expressions in Eq.~\eqref{meanvarnonpert} depend on the expectation values with respect to the initial state only, one can approximate \footnote{Indeed, the distribution is Gaussian, but only defined for positive values of $\lambda$. Thus, there will be corrections to the expectations values that can be made arbitrary small by suitably choosing the values of $\lambda_0$ and $\sigma_0^2$. To be quantitative, by setting  $\lambda_0=10$ and $\sigma_0^2=4$ as in the numerical simulations, one has a relative error at the level of the $10^{-7}$ and $10^{-5}$ level respectively for the mean and the variance.} the following terms by imposing the Gaussianity assumption: $\braket{\hat \lambda^3}_{t_0}={ \lambda}^3_{0}+3{ \lambda}_{0}\sigma_{0}^2$ and $\braket{\hat \lambda^4}_{t_0}={ \lambda}^4_{0}+6{ \lambda}^2_{0}\sigma_{0}^2+3(\sigma_{0}^2)^2$. This leads to a strong simplification of the expressions for the mean and variance, which read
\begin{subequations}
    \begin{align}\label{eq.lambda.pert.gauss}
        \braket{\hat \lambda}_{\text{Gauss},t}&=\lambda_0-2\epsilon(W_t-W_{t_0})\sigma^2_{\lambda,t_0},\\
        \label{eq.sigma.pert.gauss}\sigma^2_{\text{Gauss},t}&=\sigma^2_{0}\left[1-4\epsilon^2(t-t_0)\sigma^2_{0}\right].
    \end{align}
\end{subequations}
Now, we can directly compare these expressions with those derived from the exact, non-perturbative approach in the main text under the Gaussian assumption, which are
\begin{subequations}
    \begin{align}\label{eq.lambda.exact.gauss}
        \braket{\hat \lambda}_t&={\lambda}_{0}-2\epsilon\int_{t_0}^t \sigma^2_{\lambda,s}\D W_s,\\
        \label{eq.sigma.exact.gauss}\sigma^2_{\lambda,t}&=\frac{\sigma^2_{0}}{(1+4\epsilon^2(t-t_0)\sigma^2_{0})}.
    \end{align}
\end{subequations}
We can then straightforwardly identify Eq.~\eqref{eq.sigma.pert.gauss} as the short-time expansion of Eq.~\eqref{eq.sigma.exact.gauss}, which is valid as long as $(t-t_0)\ll1/(4\epsilon^2\sigma^2_{0})$. On the other hand, by substituting Eq.~\eqref{eq.sigma.pert.gauss} in Eq.~\eqref{eq.lambda.exact.gauss}, one obtains Eq.~\eqref{eq.lambda.pert.gauss} at the second order in $\epsilon$. A comparison of the two dynamics is shown in Fig.~\ref{fig.comparison}, but limited to times $t$ such that $\sigma^2_{\text{Gauss},t}$ remains positive.

\subsection*{State of the art on quantum-to-classical transition in cosmology}

Within quantum cosmology, the term \textit{quantum-to-classical transition} can account for two different contexts. Thus, it is essential for our discussion to clarify the differences between them. 

\textit{Quantum cosmological perturbations in a classical space-time.--} The starting point is a curved but fixed classical geometry within which fields evolve. Building on such a structure, one studies the quantum-to-classical transition of perturbations to these fields. Their origin can be quantum, but their observations are in line with a classical description.

An example of this kind of transition is that of the perturbations of the inflaton field, that can be described in terms of the Mukhanov-Sasaki variable. What we measure today are the temperature fluctuations of the CMB, which can be described as classical stochastic processes. However, within the cosmological inflation theory, such perturbations are initially of quantum origin. In this context, it was pointed out that there is not a completely satisfactory description of this transition \cite{perez2006}, and phenomenological models leading to the  collapse of the inflaton field and its perturbations were proposed as a possible solution \cite{Unanue2008,Diez2012}. It was later suggested \cite{MartinPRD2012}  that  collapse models can offer an explanation of the emergence of a single outcome of the observed CMB map. At the same time, they lead to deviations from QM that can be confronted with observations. Indeed, some consider the modifications due to the collapse dynamics to the power spectrum of the Mukhanov-Sasaki variable  \cite{MartinPRL2020,GundhiPRL2021}. The resulting corrections allow to constrain the phenomenological collapse parameters. 

Finally, we also mention that the CMB spectrum has also been motivated from a De Broglie-Bohm perspective \cite{Pinto-Neto2012}. Within this framework, one can derive a guidance equation for the Mukhanov-Sasaki variable, which differs from that of the standard cosmology by a quantum force  being negligible in the classical limit.

\textit{Emergence of a classical space-time.--} Space-time cannot be a classical entity if the gravitational field is fundamentally quantum. Therefore, there must be a mechanism to explain the classical appearance of space-time at large scales that is consistent with our observations. As in this case the quantum system under study is the whole structure of space-time, then the system is a truly closed quantum system. This means that there is nothing physically exterior to this system that could potentially lead to a collapse.

The incorporation of collapse models to modify the WDW equation was previously suggested as a solution to the problem of time \cite{Okon2014,Banerjee2015}. To the extent of our understanding, their approach consists in writing down a Schr\"odinger evolution with respect to the time coordinate $\tau$ associated with a specific space-time foliation, and to add a Hamiltonian encoding the effects of collapse models. In this way, wavefunctions which satisfy the WDW constraint will now possess a non-trivial evolution with respect to $\tau$.  However, due to the time reparameterisation invariance of GR, the time coordinate $\tau$ is still only a classical gauge degree of freedom, and thus does not fully account for the dynamics. 

In contrast, in our work, we do not focus on the problem of time but
on the quantum-to-classical transition of space-time, which has not been previously developed. We use one of the canonically conjugated variables of the WDW equation as a clock.  Thus, in our approach,  such a  clock is a physical quantity, a dynamical variable that can account for the dynamics. This approach also  opens the possibility of choosing different canonical variables as clocks. Once a clock variable is fixed, our model describes the emergence of a well-defined geometry
starting with a Universe in a superposition of different geometries.

\bibliography{main.bib}{}
\bibliographystyle{apsrev4-1}

\end{document}